# Android Inter-App Communication: Threats, Solutions, and Challenges

Jice Wang[a]  Hongqi Wu[bc]

[a] University of Chinese Academy of Sciences, Beijing, China
[b] Bureau of Veteran Cadres, the Chinese Academy of Sciences, Beijing, China
[c] School of Informatics and Computing, Indiana University Bloomington, Indiana, USA

____________________________________________________________________

## Abstract

Researchers and commercial companies have made a lot of efforts on detecting malware in Android platform. However, a recent malware threat, App collusion, makes malware detection challenging. In App collusion, two or more Apps collaborate to perform malicious actions by communicating with each other, which makes single App analysis insufficient. In this paper, we first introduce Android security mechanism and communication channels used by android Applications. Then we summarize the security vulnerabilities and potential threats introduced by App communication. Finally, we discuss state of art researches and challenges on App collusion detection.

____________________________________________________________________

## 1. Introduction

With the rapid development of mobile Internet industry, smartphone has become pervasive. According to the latest report released by IDC in 2017, the number of smartphone sales exceeded 1 billion 400 million, Google Android held 85.1% of global market share[1]. Meanwhile, Android becomes one of the most popular platform for propagation of malicious Applications. In 2017, 360 Internet Security Center has monitored 214 million malicious infections of android users, 585 thousands infections per day[2]. Researchers and commercial company have made a lot of effort on detecting malicious Applications. They developed tools using static analysis, dynamic analysis and machine learning based methods[3][4][5].

A recent malware threat, App collusion, makes malware detection more challenging. Malicious behaviors are accomplished by two or more Apps in these scenarios. Each App requests for a limited set of permissions and completes a particular component of work which do not seem dangerous to users. However, when they combine together, they have abilities to attack end users. For example, suppose there are two Apps in one's smartphone. App A have permissions to read user's contacts but not have internet permissions. App B have permissions to link internet but not contacts. Then two Apps may collaborate to steal contacts information. App communication is the key mechanism makes App collusion possible. Because App A can send user's contacts to App B, then App B can send it to Internet.

Android OS supports inter-component communication(ICC) that allows App communication across sandbox. There are multiple communication channels in Android system, such as intent, content provider, shared preference or external storage. All of these channels can be used by attackers to perform attacks, such as Broadcast theft, Activity hijacking, Service hijacking, Intent spoofing, Privilege escalation[6][7][17] and Application collusion attack[8]. In this paper, we focus on Application collusion attack. We will discuss state-of-art techniques on detecting Application attacks and point out challenges in this area. The rest of paper is organized as follows. Section 2 gives an overview of Android system and application. Section 3 summarizes the security vulnerabilities and potential threats introduced by App communication. Section 4 discuss state of art researches and challenges on App collusion detection. Finally, conclusions are present in Section 5.

## 2. Android Overview

Android is a mobile operating system developed by google, based on a modified version of the Linux kernel and other open source software[9]. The Linux kernel locates in the lowest layer of android system, while the Applications locate in the top layer.

Android Applications can be written in Kotlin, java or C++ languages. The Android SDK tools compile code along with any data and resource files into an APK, an Andorid package, which is an archive file with an .apk suffix[10].Every Android App runs in its own Linux process, and it protected by permissions and security sandbox. Each process has a unique user identifier(UID) and its own virtual machine(VM), so an App's code runs in isolation from other Apps. One android App contains four types of components: Activities, Services, Broadcast receivers and Content providers. Each type has a distinct purpose and can by communicate through Intent.

### 2.1 Android security mechanism

### 2.11 Application sandbox

The android platform uses the Linux user-based protection as a means of identifying and isolating Application resources. Each Application has a unique user ID and run it in a separate sandbox, it does not have access to the rest of the system's resource unless are granted access permissions by the user. The sandbox is simple, auditable, and based on user separation of process and file permissions[11].

### 2.12 Android permission

Android uses permission mechanism to protect the privacy of Android user. Android Application must request permission to access sensitive user data and certain system resources, such as camera and internet. An App must declare its permissions in its manifest file. Before Android 6.0, dangerous permissions are prompted at install time, while after Android 6.0, they are prompted at runtime. There are two protection levels that affect third party Apps: normal and dangerous permissions.

*Normal:* A very little risk permission that affect user's privacy or the operation of other Apps. If an App declares in its manifest that it needs a normal permission, the system automatically grants the App that permission at install time.

*Dangerous:* A high-risk permission that involve the user's private information, or could potentially affect the user's stored data or the operation of other Apps. If an App declares that it needs a dangerous permission, the user has to explicitly grant the permission to the App. To use a dangerous permission, App must prompt the user to grant permission at runtime.

*Permission groups:* Permissions are organized into groups related to a device's capabilities or features. For example, the SMS group include both the READ_SMS and the RECEIVE_SMS permissions. And for convenience, if an App has already been granted another dangerous permission in the same permission group, the system immediately grants the permission without any interaction with the user[12].

## 2.2 Inter-App communication channels

### 2.21 Intent

Intent is a messaging object which can used to request an action from App component. The three main usage are start activity, start service, and send broadcast, and there are two types of intents:

*Explicit intents:* an Application will use the intent by supplying either the target App's package name or a fully-qualified component class name. The typically use of explicit intent is to start a component in one's own App, because you know the class name of the component you want to start.

*Implicit intents:* you do not name a specific component, but instead declare a general action to perform, which allows a component from another App to handle it.

### 2.22 content provider

Content provider store and provide content to Applications like relation database. They encapsulate data and provide it to Applications through the single ContentResolver interface. A content provider is only required if you need to share data between multiple Applications[13].

### 2.23 shared preference

Shared preference is a small collection of key-values that you'd like to save, and you should use the SharedPreferences APIs. A SharedPreferences object points to a file containing key-value pairs and provides simple methods to read and write them, and Applications can use shared preference to exchange information[14].

### 2.24 external storage

Android device supports a shared "external storage" space that you can use to save files. It is a storage space that users can mount to a computer as an external storage device or physically removable. Files saved to external storage are world-reachable, so Applications can use it share information.

## 2.25 covert channel

Android covert channel is a medium which use shared resources for communication between two malicious entities without using conventional methods[15].Timing and storage channel are two types of covert channels. The timing channels need time synchronization between two colluding Apps since they do not store information. on the contrary, the storage channel need to store information in shared resource attribute which have some form of memory.

# 3. Inter-App communication related threat

## 3.1 Broadcast theft

A public broadcast sent by Application is vulnerable. A malicious App can silently read the public broadcast while the actual recipient is also listening. An active attacker could even register itself as a high priority and then receive the Intent first, so it can cancel it later and cost denial of service attacks.

## 3.2 Activity hijacking

A malicious could register an implicit Intents, and a malicious activity might be launched if it receives another Application's implicit Intents. In this situation, the malicious activity could read the data in the Intent and then immediately relay it to a legitimate activity[6].

## 3.3 Service hijacking

The service hijacking is similar as activity hijacking, it occurs when a malicious service intercepts an implicit Intent. The service hijacking is not Apparent to users and it can be used for Phishing, Denial of Service and component hijacking attacks[6][7].

## 3.4 Intent spoofing

Intent spoofing is launched by a malicious Application sending an Intent to an exported component that do not expect it. If the victim Application takes some action upon receipt of such an Intent, the attacker can trigger it[6].

## 3.5 Privilege escalation

Privilege escalation occurs when an Application with less permissions is not restricted to access components of a more privileged Application[17]. Confused deputy attacks is one kind of privilege escalation. In this scenarios, a malicious Application exploits the vulnerable interfaces of another privileged Application[18].

## 3.6 Application collusion attack

Application collusion attack is a scenario where two or more Applications collaborate to perform malicious action. Each Application has limit permissions and resources and only does part of job. Through legitimate communication channels, they can work together. And they do not need to take advantage of system vulnerabilities to perform malicious action such as privacy leakage. Because each App performs legitimate, it is really hard for malicious detection.[18][19].Recently, intra-library collusion is also proposed, which happens when individual libraries obtain greater privileges by being embedded within multiple Apps. It is similar as Application collusion expect collusion behaviors happens in third-party libraries[20].

## 4. App collusion detection techniques and challenges

In these section, we discuss state of art App collusion detection techniques, point out their limitations and future challenges.

### 4.1 detection techniques

In order to detect App collusion using Intent-based communication channel, we have following problem should be solved step by step. First, wo should o how to detect intent usage in Applications, and we should extract communication information in intent. Second, in order to detect Application Collusion, we should have ability to find inter-App communication in App sets. Next, after we detect App communication through intent channels, we need to know whether such ICC is malicious or benign. Finally, as the huge number of Apps available in the Android market, detection methods should be scalable. Different researches focus on different problems, so we summarize researches in form of problem & solutions.

***Problem1: how to analyze intent information***
***Solutions:***

Octeau D et al.[21] developed a static analysis tool, Epicc, to discover inter-component communication in android Applications. It transforms ICC analysis problem into an Interprocedural Distributive Environment(IDE) problem. The analysis is soundness and precision, also scales well. The experiment dataset is 1200 Applications selected from the Play store, the tool can identity over 93% of ICC locations for all Applications studied. After Epicc, IC3 was released, which is more precise than previously released ICC tool. Chin E et al.[22] present ComDorid to detect vulnerabilities in android Applications. It parses the disassembled code and logs potential component and Intent vulnerabilities. It performs flow-sensitive, intraprocedural static analysis, with limited interprocedural analysis that follows method invocations to a depth of one method call. Comdroid tracks the state of Intents, IntentFilters, registers, sinks, and components. For each method that uses Intents, it tracks the value of each constant, string, class, Intent, and IntentFilter.

***Problem2: how to analyzing pairs of Apps***
***Solutions:***

Currently, the number of real colluding Apps available to researchers is limit. Blasco J et al.[23] propose a system called Application Collusion Engine to automatically generate combinations of colluding and non-colluding Apps. It has create more than

5000 different App sets for research community. Klieber W et al.[24] use static taint analysis for tracking data flow in a set of android Applications. The taint flow analysis takes place in two phases. In phase 1, data flow in each Application is tracked by FlowDroid[25], identification of the properties of sent intents is found by Epicc[21]. Phase 2 use the output of phase 1 and construct the source-to-sink flows found in the set of Apps. Tsutano Y et al.[26] introduce JITANA, a program analysis framework designed to analyze multiple android Applications simultaneously. It operates similar as a classloader used in any Java or Android virtual machine. Engineers can load parts of an App, an entire App, or multiple Apps for static analysis. As it can analyze all loaded Apps simultaneously, it can generate program analysis graphs that span multiple Apps, which can be used for analyzing interacting Apps. Li L et al.[27] Developed a tool called ApkCombiner that reducing an inter-App communication problem to an intra-App inter-component communication problem. It combines different Apps into a single apk so that existing tools can indirectly perform inter-App analysis.

***Problem3: how to classification of ICC into malicious and benign***
***Solutions:***

Elish K O et al.[28] construct ICC Maps to capture communication ICC channels of 2644 real benign Apps. They find existing permission-based collusion-detection policies trigger a large number of false alerts in benign, which shows the need on classifying benign and malicious ICC channels.Xu K et al.[29] propose a malware detection method named ICCDetector, which can detect stealthy collusion attack. ICCDetector trains a set of benign Apps and a set of malwares, outputs a detection model, and then employs the trained model for malware detection. ICCDetector also provides a systemic analysis of ICC patterns of benign Apps and malwares.Lee Y K et al.[30] present SEALANT, a technique that use static analysis to infer vulnerable communication channels , with runtime monitoring of these channels and helps to prevent attacks. It only focus on following types of vulnerabilities, intent spoofing, unauthorized intent receipt and privilege escalation, so it can define malicious behaviors and design algorithm to find vulnerable ICC paths.

***Problem4: how to detect App collusion in large-scale.***
***Solutions:***

Liu F et al.[31] present MR-Droid, they use MapReduce-based computing framework to perform accurate and scaladable inter-App ICC analysis. Based on MapReduce, MR-Droid can quickly processing large-scale app-pairs, which makes detection of App collusion much quicker.Sattler F et al. [32] presented a variability-aware approach to inter-app data-flow analysis. It use a graph-based data structure which represent inter-App flows , the structure exploits redundancies among flows and thereby prevents the combinatorial explosion.Octeau D et al.[33] introduce a probabilistic model of ICC on top of static analysis results, they design an algorithm for performing link resolution and triage them using the probabilistic approach. They find over 95.1% of potential links are associated with probability values below 0.01 and are thus likely unfeasible links, which make it possible to consider only a small subset of all links without significant loss of information. And it makes inter-application analysis more scalable.

## 4.2 Challenges

Although researchers have proposed different kinds of techniques for inter-App analysis, but they still have some limitations. Most of current research using static analysis, they can not handle java reflection code obfuscation and reinforcement Application. Also, current static analysis does not support analyze native code which might be used by quite number of malwares. As for dynamic analysis, they only concentrate on certain types of App collusion attacks, which limit its usage. Also, dynamic analysis consumes much more time than static analysis, and lacks of scalability. Precision and scalability are still big challenges which should be solved.

Thera are two types of inter-App communication channels: Intent-based communication channel and covert channel. Current researches have focused on detection App collusion through Intent-based communication channels. To the best of our knowledge, there is no research aim to detect App collusion through Covert channels. Researchers should pay more attention to this area in the future.

# 5. Conclusion

App collusion is an increasing threat in Android platform. This paper concludes state of art research progress in detection of App collusion. In this paper, we introduce Android system and its security mechanism, then we conclude the communication channels used by android Applications. Next, we summarize the security vulnerabilities and potential threats introduced by App communication. Finally, we discuss state of art researches on App collusion detection and point out challenges in this area. This challenges must be solved and that require further researches.